\documentstyle[12pt]{article}
\title{Coordinate-free derivation of Yang-Mills-Chern-Simons field equations}
\author{Marcos Jardim \thanks{This work is supported by a grant from 
CNPQ-Brazil} \\ State University of Campinas \\ 
Departament of Mathematics \\ e-mail:{\em jardim@ime.unicamp.br}}
\date{February 96}

\begin{document} \maketitle 

\begin{abstract}
In this paper we derive in a coordinate-free manner the field equations 
for a lagrangean consisting of Yang-Mills kinetical term plus Chern-Simons
self-coupling term.  This equation turns out to be an eigenvalue equation for
the covariant laplacian.
\end{abstract} \pagebreak

\section{Introduction}
Three-dimensional topologically massive gauge theory was introduced by 
Deser, Jackiw \& Templeton \cite{DJT} and Schonfeld \cite{S} in early 80's 
and has been object of study by many authors since then. In the abelian
case, eletric charged particles are provived with fractional statistics and
the theory may be applied to the study of the quantum Hall effect and the 
high temperature superconductors.

The mathematical setting is a principal fibre bundle {\bf P} over a 
three-dimensional {\bf M} and structural group {\bf SU}(n), provided with
a connection $A$ whose curvature is $F=dA+A\wedge A$. As is weel-known, $A$
and $F$ play the role of the gauge potential and the gauge field strength,
respectivelly. We recall that $A$ is a su(2)-valued 1-form in {\bf M}, and
$F$is a su(2)-valued 2-form. The lagrangean under consideration is, in its
coordinate-free form, the following:
\begin{equation} \label{ymcs}
{\cal L}=\frac{1}{2g^{2}}{\rm tr}\left\{F\wedge*F-
\mu\left(A\wedge dA+\frac{2}{3}A\wedge A\wedge A\right)\right\}
\end{equation}
where $g$ in the gauge coupling constant with units [mass]$^{-1/2}$ and
$\mu$ is a new constant with units [mass]; this constant will be seen to
be the topological mass of the gauge field. This name comes from the fact
that the Chern-Simons term is a {\em topological} term, meaning metric 
independence. 

The first thing to note is that the second term (the Chern-Simons term) in 
\ref{ymcs} is not gauge invariant, but behaves in a very special manner under 
gauge transformations. In fact, the Chern-Simons term can be identified with 
the second secondary characteristic class $TC_{2}(A)$ defined by Chern \& Simons
in \cite{CS}; theorem 3.16 of this work asserts that when the second Chern 
class of {\bf P} is zero (and this is indeed the case, because every 
SU(2)-bundle over three-manifols is trivial) then $TC_{2}(A)$ defines a 
{\bf R}/{\bf Z}-cohomology class in {\bf M}. So, under gauge transformations
the action ${\cal S}=\int_{M}{\cal L}$ changes by the addition of an integer
multiple of $\mu$. Thus, for the phisically meaningful quantity 
$\exp(i{\cal S})$ be gauge invariant the constant must be quantized 
$\mu=2\pi n$. This picture is slightly diferent if the gauge group is U(1);
in this case, $TC_{2}(A)\equiv 0$, thus ${\cal S}=\int_{M}{\cal L}$ is gauge
invariant and there is no quantization condition in the eletromagnetic case.

\section{Derivation of the field equation}
We proceed to the main objective of this letter, that is the coordinate-free
derivation of the equations of motion of the lagrangean given by \ref{ymcs}.
Introducing a first order variation on the gauge potential 
$A_{\tau}=A+\tau a$, which effect on the gauge field is given by 
$F_{\tau}=F+\tau Da+\tau^{2}a\wedge a$ ($D$ denotes exterior covariant 
differentiation), we get:
$$ \begin{array}{rcl}
{\cal L}_{\tau}&=&(F+\tau Da+\tau^{2}a\wedge a)\wedge*(F+\tau Da+\tau^{2}a\wedge a)+ \\
&&\!\!\! \mu\left((A+\tau a)\wedge(dA+\tau da)+\frac{2}{3}
(A+\tau a)\wedge(A+\tau a)\wedge(A+\tau a)\right) \\ 
&=& {\cal L}+\tau\left\{
F\wedge*Da+Da\wedge*F+
\mu\left(a\wedge dA+A\wedge da+\right.\right.\\ &&\left.\left.
\frac{2}{3}(a\wedge A\wedge A+A\wedge a\wedge A+A\wedge A\wedge a)\right)
\right\}+O(\tau^{2})
\end{array} $$
and asking that $\frac{d}{d\tau}{\cal L}_{\tau}|_{\tau=0}=0$, we have:
\begin{equation} \label{xxx} \begin{array}{l} 
F\wedge*Da+Da\wedge*F+ \\
+\mu\left[a\wedge dA+A\wedge da+\frac{2}{3}(a\wedge A\wedge A+A\wedge a\wedge A+
A\wedge A\wedge a)\right]=0 
\end{array} \end{equation}
We can express \ref{xxx} in a better way. Recall that the space of $k$-forms
can be given the following inner product, where $\psi_{1}$ and $\psi_{1}$
are $k$-forms and $*$ is the Hodge star operator:
\begin{equation} \label{yy}
\left<\psi_{1}|\psi_{2}\right>=\int_{M}tr(\psi_{1}\wedge*\psi_{1}) 
\end{equation}
Note also that $**=1$ for 1- and 2-forms in 3-manifolds and that $*$ is a
self-adjoint operator. Thus:
$$ \begin{array}{l}
2\left<Da,F\right>-\mu\left[\left<a,*dA\right>+\left<A,*da\right>+\right. \\
\left.\frac{2}{3}(\left<A\wedge A,*a\right>+\left<A\wedge a,*A\right>
+\left<a,*(A\wedge A)\right>)\right]=0
\end{array} $$
and being $D^{*}=*D*$ and $\delta=*d*$ are the formal adjoints of $D$ and 
$d$, respectivelly, we get:
$$ 2\left<Da,F\right>-\mu\left[\left<a,*dA+\delta*A\right>+
\frac{2}{3}(2\left<a,*(A\wedge A)\right>+\left<*(A\wedge a),A\right>)\right]=$$
$$ =2\left<a,D^{*}F\right>-\mu\left[2\left<a,*dA\right>+
2\left<a,*(A\wedge A)\right>\right]=0 $$
because $\delta*A=(*d*)*A=*d(**A)=*dA$ and 
$\left<*(A\wedge a),A\right>=A\wedge**(A\wedge a)=**(A\wedge A)\wedge a=
\left<a,*(A\wedge A)\right>$, so:
$$ \left<a,D^{*}F\right>-\mu\left[\left<a,*(dA+A\wedge A)\right>\right]= 
\left<a,D^{*}F-\mu(*F)\right>=0 $$ 
for any su(2)-valued 1-form $a$. Finally, we deduce that the field equation
is given by:
\begin{equation} \label{eqmov}
D^{*}F-\mu(*F)=0
\end{equation}

Observing that $*F=(*F)_{\alpha}=\epsilon^{\mu\nu\alpha}F_{\mu\nu}$ and
that $D^{*}F=-2D_{\mu}F^{\mu\nu}$ we get the field equation \ref{eqmov}
in its usual coordinate form:
$$ D_{\nu}F^{\nu\alpha}+\frac{\mu}{2}\epsilon^{\beta\nu\alpha}F_{\beta\nu}=0 $$

It is easy to see that \ref{eqmov} is gauge invariant, although the 
lagrangean is not.

 \end{document}